\documentclass{article}

\usepackage{arxiv}
\usepackage{amsmath, amsthm, amssymb, amsfonts, braket}
\usepackage[utf8]{inputenc} 
\usepackage[T1]{fontenc}    
\usepackage{hyperref}       
\usepackage{url}            
\usepackage{booktabs}       
\usepackage{amsfonts}       
\usepackage{nicefrac}       
\usepackage{microtype}      
\usepackage{lipsum}
\usepackage{graphicx}
\usepackage{xcolor}
\graphicspath{ {./images/} }

\usepackage[numbers,square]{natbib}
\title{Non-Markovian Quantum Control via Model Maximum Likelihood Estimation and Reinforcement learning}

\author{
 Tanmay Neema \\
  Computer Science Laboratory\\
  SRI International\\
  Menlo Park, CA 94025 \\
  \texttt{tanmay.neema@sri.com} \\
  \And
 Susmit Jha \\
  Computer Science Laboratory\\
  SRI International\\
  Menlo Park, CA 94025 \\
  \texttt{susmit.jha@sri.com} \\
   \And
 Tuhin Sahai \\
  Applied Sciences\\
  SRI International\\
  Menlo Park, CA 94025 \\
  \texttt{tuhin.sahai@sri.com} \\
}

\begin{document}
\maketitle
\begin{abstract}

Reinforcement Learning (RL) techniques have been increasingly applied in optimizing control systems. However, their application in quantum systems is hampered by the challenge of performing closed-loop control due to the difficulty in measuring these systems. This often leads to reliance on assumed models, introducing model bias, a problem that is exacerbated in open quantum dynamics where Markovian approximations are not valid. To address these challenges, we propose a novel approach that incorporates the non-Markovian nature of the environment into a low-dimensional effective reservoir. By initially employing a series of measurements as a 'dataset', we utilize machine learning techniques to learn the effective quantum dynamics more efficiently than traditional tomographic methods. Our methodology aims to demonstrates that by integrating reinforcement learning with model learning, it is possible to devise control policies and models that can counteract decoherence in a spin-boson system. This approach may not only mitigates the issues of model bias but also provides a more accurate representation of quantum dynamics, paving the way for more effective quantum control strategies.

\end{abstract}

\section{Introduction}

Quantum computation is an exciting area with the potential to revolutionize material and drug discovery~\cite{cao2018potential}, cryptography~\cite{rieffel2000introduction}, optimization and search~\cite{marsh2020combinatorial}, and efficient simulation of high-dimensional dynamical systems~\cite{q_sde,surana2022carleman}. By exploiting quantum mechanical phenomena such as superposition and entanglement, researchers have constructed novel algorithms that provide exponential to quadratic speed-up over state-of-the-art approaches. For example, quantum algorithms provide exponential gains for the task of computing prime factors for any given integer~\cite{shor1999polynomial}. However, all these gains require the development of fault tolerant quantum platforms that are robust to external noise.

Quantum control and state initialization remain central problems towards the development of fault tolerant and robust quantum computing as according to DiVincenzo's criteria~\cite{georgescu2020divincenzo}.
Generally, current methods of quantum control rely on either assumed models and their accessible gradients
and/or rely on feedback through measurements~\cite{wu2018data,sivak2022model, Morzhin2018, Dong2009, Chen2021, Dong2009SMC} .
For example, Krotov's algorithm minimizes a functional $J$ that represents the update cost of the control parameters $\{ \epsilon^i_l(t) \}$ at the i-th iteration and the infidelity $J_T$ of the final propagated state and target state~\cite{yu2021many, Morzhin2018}. In another example of the Gradient Pulse Engineering algorithm, the fidelity is maximized across the time discretized control parameters through gradient ascent ~\cite{grape2}. Notably, these methods assume model knowledge in finding analytical gradient. As a result, optimal control theory may not be robust to open quantum dynamics where additional dissipative and decohering terms remain unmodeled. To illustrate this point, many of these models work utilize Master equations at the limit of the Markovian approximation which assume that the coupling to the environment is weak or the inital state is uncorrelated with the environment ~\cite{Lidar2019, Breuer2007}, which for the purpose of state initialization is an unreasonable assumption. Consequently, these control policies fail to produce desirable final state fidelities when transfering from simulation to experiment.

In regards to feedback-based control schemes, the process of measurement of a quantum system extends duration of control process dependent on interaction times with the measurement device, often acting as computational bottlenecks. Strasberg et al. demonstrated that time of measurements scale with the dimensionality of the system \cite{Strasberg2020}.  In addition, the process of measurement, both strong and weak measurements measurements, necessarily collapse the system into an eigenstate for a projection-valued measures (PVM) or an unknown state for positive operator valued measures (POVM),  where the probability of a measurement depends on the trace of the state and the measurement operator \cite{Beneduci2020}). As a result, feedback dependent methods have undefined protocol lengths, dependent on the inherent stochasticity of a quantum measurement such as in a sliding mode controller described in the work byDong et al.  \cite{Dong2009SMC}. Our proposed method attempts to mitigate both of these issues, through a model learning phase and policy learning phase.

Reinforcement learning has been demonstrated to be successful in quantum control applications ~\cite{Niu2019, Zhang2019, Ma2020, He2021, Wei2019, Bukov2017}. Zhang et al. compared the performance between Krotov's, stochastic gradient descent (SGD), tabular Q learning (TQL), deep Q learning (DQL), and policy gradient(PG) and found that DQL and PG performed best ~\cite{Zhang2019}. These RL methods led to more simplified control sequences, leading to shorter operation times and  higher fidelities. Niu et al. demonstrated that trusted-region policy optimization via reinforcement learning was robust to leakage and stochastic gate errors ~\cite{Niu2019}. Ma et al. employed curriculum-based reinforcement learning and demonstrated that it was superior to gradient and genetic algorithms, even in the increase in the dimensionality of the system and was capable of perform the protocol in fewer steps ~\cite{Ma2020}. The robustness of the reinforcement learning in the presence of noise, its effectiveness in larger systems, and the protocal speed ups motivate the usage of RL in our proposed method.

Still these methods suffer from model-bias, which can lend to poor application to experimental setups. As a result, as a part of the model learning phase, our proposed method aims to learn the open quantum dynamics of the system from measurements. 

For a general dynamical map, the quantum process tomography grows exponentially with the system size and relies on enough measurements to find an accurate reconstruction, a problem further exacerbated by the fact that the evolution superoperators do not form a semigroup. In general, the dynamics are described by a memory kernel, $\mathcal{K}$,  in the Nakajima Zwanzig equation ~\cite{Lidar2019}. $\frac{d}{dt}\rho_S(t) = -iL_s\rho_S(t) + \int_0^t d\tau \, \mathcal{K}(t-\tau) \rho_S(\tau)$

Finding the exact form of the kernel can be accomplished through determining a series of transfer tensors ~\cite{Cerrillo2013, Gherardini2021, Pollock2017, Chen2019}. The evolution is described by,
\begin{equation} 
\rho(t_m) = 
\begin{pmatrix}
T_1 & T_2 & \cdots & T_K
\end{pmatrix}
\begin{pmatrix}
\rho(t_{m-1}) \\
\rho(t_{m-2}) \\
\vdots \\
\rho(t_{m-K})
\end{pmatrix},
\end{equation} 

where ${T_i}$ can be found as a function of the dynamical maps between any time $t_0$ to $t_i$ and is truncated at $T_K$ as the long-term correlations approach 0. This too, however in practice, encounters the same difficulty of quantum tomography methods, i.e., having to learn the series of dynamical maps. In an attempt to reduce sampling size Torlai et al. recently utilized unsupervised learning ~\cite{Torlai2023}, indicating that machine learning has utility in generalizing quantum dynamics. 

 Recent work by Luchnikov et al. ~\cite{Luchnikov2019} demonstrated that with a sufficiently long sequence of measurements of a quantum system, non-Markovian effects, even in strongly coupled regimes, may be learned, demonstrating a sharp distinction between traditional tomographic approaches where a series of identical states are prepared and evolved. Here, by evolving the system with intermediate and equally spaced projective measurements, the model attempts to maximize the probability a sequence of measurements $p(\{E_i\} | \mathcal{L}_{S+ER})$ over an iteratively learned generator $\mathcal{L}_{S+ER}$ that describes the evolution of the system and environment embedding. Luchnikov et al. found that the method was able to generate maps with a given error $\epsilon$ in $\epsilon^{-2}$ samples independent of the specific map. In traditional tomographic methods, one also needs $\epsilon^{-2}$ samples  for given error $\epsilon$, they require $K$ times more measurements to produce $K$ maps with the same error (e.g. $\epsilon \propto \sqrt{\frac{K}{n}}$). 

Treating the model learning as overhead, the proposed method simulates the quantum system approximately and utilizes reinforcement learning agent as the controller. Notably, we demonstrate that that the suggested Markovian embedding can still be useful in highly decohered system such as in a relaxed spin boson, i.e., in highly correlated systems and environments.  In addition, we demonstrate that a reinforcement learning agent can actively bring the system away from thermal equilibrium through controls that apply directly to the system.

The following paper is structured as follows, section 2 describes the simulated model of interest and the dynamics of the two state system. Section 3 details the learning of the open dynamics and details the derivation of the iterative learning of the quantum map that maximizes the probability of a sequence of the simulated data. Section 4 introduces the actor-critic policy gradient reinforcement learning algorithm and the choice of reward function. Finally, Section 5 concludes by describing the results and discusses future implications.

\section{Model of Open Quantum System}
\label{sec:headings}

For two state systems, such as in qubits, spin-boson model has been able to effectively capture the environment and system interactions. The model consists of a two state system and its linear interaction with a spectrum of harmonic oscillators. Inspired by this model, we simulate a system with the following Hamiltonian,

\begin{equation}
H_{\text{total}} = H_0 + H_c(t) + H_{\text{bath}} + H_{\text{int}},
\end{equation}

where each individual Hamiltonian term are given by:

\begin{align}
H_0 &= -\frac{g}{2} B_0 \sigma_z, \\
H_c(t) &= -\frac{g}{2} B_x(t) \sigma_x, \\
H_{\text{bath}} &= \sum_k \hbar \omega_k \hat{a}_k^\dagger \hat{a}_k, \\
H_{\text{int}} &= \sigma_z \sum_k \lambda_k (\hat{a}_k^\dagger + \hat{a}_k).
\end{align}

Here, \( H_0 \) represents the static part of the Hamiltonian, \( H_c(t) \) the control Hamiltonian which may be time-dependent, \( H_{\text{bath}} \) the Hamiltonian for the bath of oscillators (which models the environment) with a discrete range of frequencies that is characterized by the spectral density $J(\omega) = \sum_k |\lambda_k|^2 \delta(\omega - \omega_k)$, and \( H_{\text{int}} \) the interaction Hamiltonian between the qubit and the bath. 

For simplicity, we assume that the environmental noise contains a single mode, i.e., the qubit is coupled to a single quantum harmonic oscillator (QHO),  with $J(\omega) = \lambda_0^2 \delta(\omega - \omega_0) $ and $\omega_0 \approx \frac{gB_0}{2 \hbar}$ so that the QHO frequency is near resonant with the qubit system, ensuring stronger non-Markovian interactions.

During model learning, we initialize the simulated bath state from thermal equilibrium at a temperature $T$ such that the energy of the oscillator can be described by the equipartion theorem i.e. $k_B T \gg \hbar \omega_0 $ (we set $k_B = \hbar = 1$). So for the simulated bipartite qubit environment system, this would be equivalent to $\rho(0) = \rho_S(0) \otimes \rho_{bath}(0) =  \ket{e}\bra{e} \otimes \frac{1}{Z} e^{\frac{-H_{bath}}{T}}$. 

The control policy being able to navigate the control landscape of the open quantum system utilizing interactions exclusively with the system (which would be unable to otherwise due to unitary nature of these transformations) would indicate a deep understanding of the dynamics gleaned from the model and would indicate an ability to reverse relaxation and dissipative activity.

\section{Learning Open Quantum Dynamics}

In this section, we describe the model learning algorithm which follows the description provided by Luchnikov et. al.~\cite{Luchnikov2019}. We describe an equivalent but different derivation for greater clarity. 

The dynamics of the open quantum system is broken into two parts, an effective reservoir (ER) and the system (S) and the method attempts to learn the dynamics of the reservoir interactions through a series of measurements performed on the system of interest. A dataset of projective value measurements $\{E_i\}$ at each time steps $t_i$, i.e., the system of interest (in our case simulation) is evolved under open quantum dynamics for time $\Delta t$ and then is followed by a measurement within an orthonormal basis $\{\ket{\psi_k} \bra{\psi_k}\}$. $E_i$ is the measurement from the set $\{\ket{\psi_k}\bra{\psi_k}\}$ associated with an outcome $k$ at time step $t_i$ with a probability $p_i = \text{tr}(\ket{\psi_k}\bra{\psi_k} \rho_S(t_i) \ket{\psi_k}\bra{\psi_k}$. The total probability of the sequence of $\{E_i\}$, $p = \Pi_{i=1}^n p_i$. As $n \rightarrow \infty$, an expectation would be that probability of said series of measurements would be maximized under the real system dynamics. Motivated by this intuition, we aim to learn a generator $\mathcal{L}_{S+ER}$ that maximizes the probability $p(\{E_i\} |  \mathcal{L})$, 

 $$\mathcal{L}^* = \underset{\mathcal{L}}{\arg\max} \, p. $$

Evidently, we have no means to directly measure the environment, hence, as a hyperparameter, we vary the estimated dimensionality of the system $d_{ER}$. We consider the evolution of a $\rho_{S+ER}$ which once traced out yields the approximate system dynamics of $\rho_S$ given by,

\begin{equation}
\dot{\rho}_{S+ER} = \mathcal{L}\rho_{S+ER}.
\end{equation}

The channel representing the evolution between each time step is given as $\Phi$,
\begin{equation}
\Phi = e^{\mathcal{L}_{S+ER} \Delta t}.
\end{equation}

Given the channel $\Phi$, we now express the probability of  ${E_1, E_2, E_3 ...}$ spaced at times $\Delta t$ as:
\begin{equation}
p = \text{tr}\left(E_n \Phi\left[E_{n-1} \Phi\left[\ldots \left(E_1 \Phi\left[\rho_0\right] E_1\right) \ldots\right] E_{n-1}\right] E_n\right)
\end{equation}

An equivalent expression of $p$ can be given by the forward and backpropagating operators $\Tilde{\rho_i}$ and $\mathcal{E}_i$ respectively,

\begin{equation}
p = \text{tr}\left(\Tilde{\rho_i} \mathcal{E}_i\right) \quad \forall i \in \{0, 1, 2, \ldots, n\},
\end{equation}
where:
\begin{align}
\Tilde{\rho_i} &= E_{i} \Phi[\Tilde{\rho}_{i-1}] E_{i}, \quad \text{with} \quad \Tilde{\rho_0} = \rho_0, \\
\mathcal{E}_i &= \Phi^{\dagger}\left[E_i \mathcal{E}_{i+1} E_i\right], \quad \text{with} \quad E_n = I.
\end{align}

Na\"ively, one could initialize $\Phi$ with its ${(d_Sd_{ER})}^4$ parameters and apply an optimization algorithm to learn them, however, since $\Phi : B(\mathcal{H}_{S+ER}) \rightarrow B(\mathcal{H}_{S+ER})$ ($\mathcal{B}$ being the set of bounded linear operators) satisfies the properties of a valid quantum channel, the set of parameters is constrained.

A valid quantum channel is completely positive and trace-preserving (CPTP). According to the Stinespring Dilation Theorem, for a CPTP map $\Phi$, there exists a representation:
\begin{equation}
\Phi[\rho_{S+ER}] = \text{tr}_A(U (\rho_{S+ER} \otimes \rho_A) U^{\dagger})
\end{equation}
with $U = e^{-iH\Delta t}$, $H \in B(\mathcal{H}_{S+ER}\otimes \mathcal{H}_{A} )$, and $\rho_a$ is a fixed ancilla system. 

The statement of the theorem implies a quantum channel can be represented by a unitary evolution of a lifted quantum system. Ideally, the dimension of $d_A = d_Sd_{ER}$ so that the set of learnable parameters still matches the dimensionality of the channel $(d_Sd_{ER})^4$.

Finding $\mathcal{L}_{S+ER}$ now comes down to learning the joint Hamiltonian of the system, effective reservoir, and ancilla, given by:

\begin{equation}
H^* = \underset{H}{\arg\max} \, p
\end{equation}

To determine this $H^*$, we need to find an analytical gradient of the probability with respect to the Hamiltonian matrix entries. The derivation is as follows,

\begin{align*}
\frac{d\log p}{dH_{\mu\nu}} 
&= \frac{1}{p} \text{tr}\left( \frac{d\Tilde{\rho}_n}{dH_{\mu\nu}}\mathcal{E}_n \right), \\
&= \frac{1}{p} \text{tr}\left( \frac{d\Phi}{dH_{\mu\nu}}[\Tilde{\rho}_{n-1}] E_n \mathcal{E}_n E_n + \frac{d\Tilde{\rho}_{n-1}}{dH_{\mu\nu}} \mathcal{E}_{n-1} \right), \\
&= \frac{1}{p} \sum_{i=0}^{n-1} \text{tr}\left( \frac{d\Phi}{dH_{\mu\nu}}[\Tilde{\rho}_{i}] E_i \mathcal{E}_i E_i \right), \\
&= \frac{1}{p} \sum_{i=0}^{n-1} \text{tr}\left( \left(\frac{dU}{dH_{\mu\nu}} \left[\Tilde{\rho}_{i} \otimes \rho_A\right] U^{\dagger} + U \left[\Tilde{\rho}_{i} \otimes \rho_A\right] \frac{dU^{\dagger}}{dH_{\mu\nu}} \right) \left( E_i \mathcal{E}_i E_i \right) \otimes I_A \right).\\
\end{align*}

 Since $p$ is function of a series of linear operators, taking the derivative can be expanded easily by the product rule, yielding the form shown. We unravel the expression of the $p$ backwards from $n$ to 0 to achieve the result. The rest follows trivially. 

 Next, we find the derivative of the unitary with respect to approximated Hamiltonian. We perform a first order expansion in $H_{\mu\nu}$ with a paramter $\epsilon$ at $\epsilon = 0$ which is equivalent to taking the derivative with the matrix element.

\begin{align*}
\frac{dU}{dH_{\mu\nu}} &= \frac{d}{d\epsilon} U(H + \epsilon \ket{\mu} \bra{\nu})\big|_{\epsilon=0} \\
&= \frac{d}{d\epsilon} e^{-i(H + \epsilon \ket{\mu} \bra{\nu})\Delta t} \\
&= \frac{d}{d\epsilon} \sum^{\infty}_{n=0} \frac{(-i\Delta t)^n}{n!} (H + \epsilon \ket{\mu} \bra{\nu})^n\big|_{\epsilon = 0} \\
&= \frac{d}{d\epsilon} \left(I + -i\Delta t H + \sum^{\infty}_{n=1} \frac{(-i\Delta t)^n \epsilon}{n!} \sum^{n-1}_{j=0} H^{n-1-j} \ket{\mu} \bra{\nu} H^{j} + \mathcal{O}(\epsilon^2) \right) \big|_{\epsilon = 0} \\
&= \sum^{\infty}_{n=1} \frac{(-i\Delta t)^n}{n!} \sum^{n-1}_{j=0} H^{n-1-j} \ket{\mu} \bra{\nu} H^{j}
\end{align*}

Since $H$ is hermitian, it can be spectrally decomposed into:

\[
H = \sum \lambda_k P_k,
\]

where $P_k P_l = \delta_{kl}P_l$ and are projectors for the eigenstates, i.e.,

\[
H^n = \sum \lambda_k^n P_k.
\]

Hence,

\begin{align*}
\frac{dU}{dH_{\mu\nu}} &= \sum_{k,l}\sum^{\infty}_{n=1} \frac{(-i\Delta t)^n}{n!} \sum^{n-1}_{j=0}  \lambda_k^{n-1-j} \lambda_l^{j}P_k \ket{\mu} \bra{\nu} P_l, \\
&= \sum_{k,l}\sum^{\infty}_{n=1} \frac{(-i\Delta t)^n}{n!}   \frac{\lambda_k^{n-1}\left(1 - \left(\frac{\lambda_l}{\lambda_k}\right)^n \right)}{1-\left(\frac{\lambda_l}{\lambda_k}\right)}P_k \ket{\mu} \bra{\nu} P_l, \\
&= \sum_{k,l}\sum^{\infty}_{n=1} \frac{(-i\Delta t)^n}{n!}   \frac{\lambda_k^{n} - \lambda_l^{n}}{\lambda_k - \lambda_l}P_k \ket{\mu} \bra{\nu} P_l, \\
&= \sum_{k,l}   \frac{ \left(e^{-i\lambda_k\Delta t} - 1 \right) - \left(e^{-i\lambda_l\Delta t} - 1 \right)}{\lambda_k - \lambda_l}P_k \ket{\mu} \bra{\nu} P_l, \\
&= \sum_{k,l}   \frac{ e^{-i\lambda_k\Delta t}  - e^{-i\lambda_l\Delta t}}{\lambda_k - \lambda_l} \braket{\psi_k|\mu} \braket{\nu | \psi_l} \ket{\psi_k}\bra{\psi_l}.
\end{align*}

Note that for $k = l$, the coefficient reduces to $-i\Delta t e^{ -i\Delta t\lambda_l}$.

Now that we have the form of the gradient of the total logarithmic probability, we use optimization methods to compute the maximum value of $\text{log}(p)$. Specifically, we use the ADAM optimizer~\cite{kingma2014adam} over the matrix entries. The ADAM optimizer adaptively estimates first and second order  moments and has low memory requirements and is appropriate since $p$ is highly non-convex.

We finish the training when the geometric mean of $\{p_i\}$,  $p^\frac{1}{n}$ stabilizes or when the algorithm has exceeded a prescribed number of training steps.

Once we have computed $H^*$, we can easily determine the channel $\Phi$ and generator $\mathcal{L^*}$. For computational ease, we treat the channel and generators as matrices over an operator basis in a Liouville space. These matrices admit the following property, 

\begin{equation}
vec(\Phi \rho) = M(\Phi) vec(\rho).
\end{equation}

Consequently, this leads to the representation of $\mathcal{L}_{S+ER}$, which can be computed by diagonalization,
\begin{equation}
M(\mathcal{L}_{S+ER}) = \frac{1}{\Delta t} \ln(M(\Phi)).
\end{equation}

\section{Actor Critic RL Controller}

Reinforcement learning is challenging in when the state space is continuous and the actor-critic method has emerged as a promising approach in these settings. Combining the REINFORCE algorithm~\cite{Sutton1998} of policy gradient and temporal difference learning, the method iteratively updates the policy and the estimated state value function. The approach tends to be sample efficient, while mitigating the high bias of value-based methods and high variance of exclusively policy-based approaches~\citep{Wang2022, Sutton1998}. 

We initialize an artificial neural network as our policy $\pi$ and state value approximation $V$, parameterized by $\theta$ and $w$ respectively. The steps in our algorithm are outlined below,

At every time step, we measure the temporal difference (TD) error (without a discount) as, 

$$\delta_t = r_t + V(s_{t+1}) - V(s_{t}),$$

where a positive $\delta_t$ implies that the state $s_t$ is in fact more valuable than the estimate. In other words, the transition reward being greater than approximated on that instant. Consequently, we update the $w$ parameters in an attempt to minimize the sum of $\delta_t^2$ through bootstrapping,

$$w := w + \alpha_w \delta_t \nabla_w V(S).$$

Similarly, we would want to minimize action probabilities that would produce these TD errors with a good measure of loss given by $-log(\pi(a_t | s_t))\delta_t$, thus,

$$\theta := \theta - \alpha_\theta \delta_t \nabla_\theta log(\pi(a_t | s_t)).$$

Concretely speaking, the actor update is inspired by the Policy Gradient Theorem with a baseline, but in contrast due to having access to an estimate of the value function, the algorithm can estimate $R_t - b(s_t) \approx \delta_t$, with $R_t$ being the cumulative reward of the subsequent states. This means that the actor-critic method is online (independent of a future state), has less variance, and converges faster, as updates are not performed with respect to that expectation estimate at the episode termination as in the REINFORCE algorithm. 

In our control problem, we define the reward at time $t$, $r_t$, as follows,

\[
r_t = F(\text{tr}_{ER}(\rho_{S+ER}(t)), \rho^*)\mathcal{D}(\rho_{S+ER}(t)) - 1,
\]

where $F$ is the fidelity of the state $\text{tr}_{ER}(\rho_{S+ER}, \rho^*)$ given by,

$$F(\rho_S, \rho_*) = tr \sqrt{\sqrt{\rho^*}\rho_{S}\sqrt{\rho^*}}.$$

We define a metric $\mathcal{D}$ as,

\begin{equation}
\mathcal{D} =F(\rho_{S+ER}, \text{tr}_{ER}(\rho_{S+ER}) \otimes \text{tr}_{S}(\rho_{S+ER})). \end{equation}  

Here the metric $\mathcal{D}$ captures the measure of separability of the environment and system. Consequently, its value is equal to 1 in the case of factorable state. Multiplying the fidelity and `separability', the system is guided to the desired final state while also being uncorrelated with the environment. 

This reward structure provides non-sparse learning signals by ensuring that a reward (necessarily negative) is given at each time step based on the product of $F(\rho_{S+ER}, \rho^*)$ and $\mathcal{D}(\rho_{S+ER})$ minus 1, which reaches a maximum of 0, when it satisfies the fidelity of 1 and 'separability' of 1. The episode is terminated at time $T$ or when $r_t > \alpha$. The termination criteria and necessarily negative rewards encourage the policy to achieve an adequate level of fidelity as quickly as possible, thus, promoting efficiency in learning. The agent learning can be thought as being divided as an exploration phase where the agent learns the system dynamics and the control effects, with the episode terminating after time $T$. This is followed by the exploitation phase where the agent acts greedily (achieving higher reward under shorter periods of time) with the episode terminating after reaching the latter condition. 

\section{Results and Discussion}

 \begin{figure}[ht]
  \centering
  \begin{minipage}{0.45\textwidth}
    \centering
    \includegraphics[width=\linewidth]{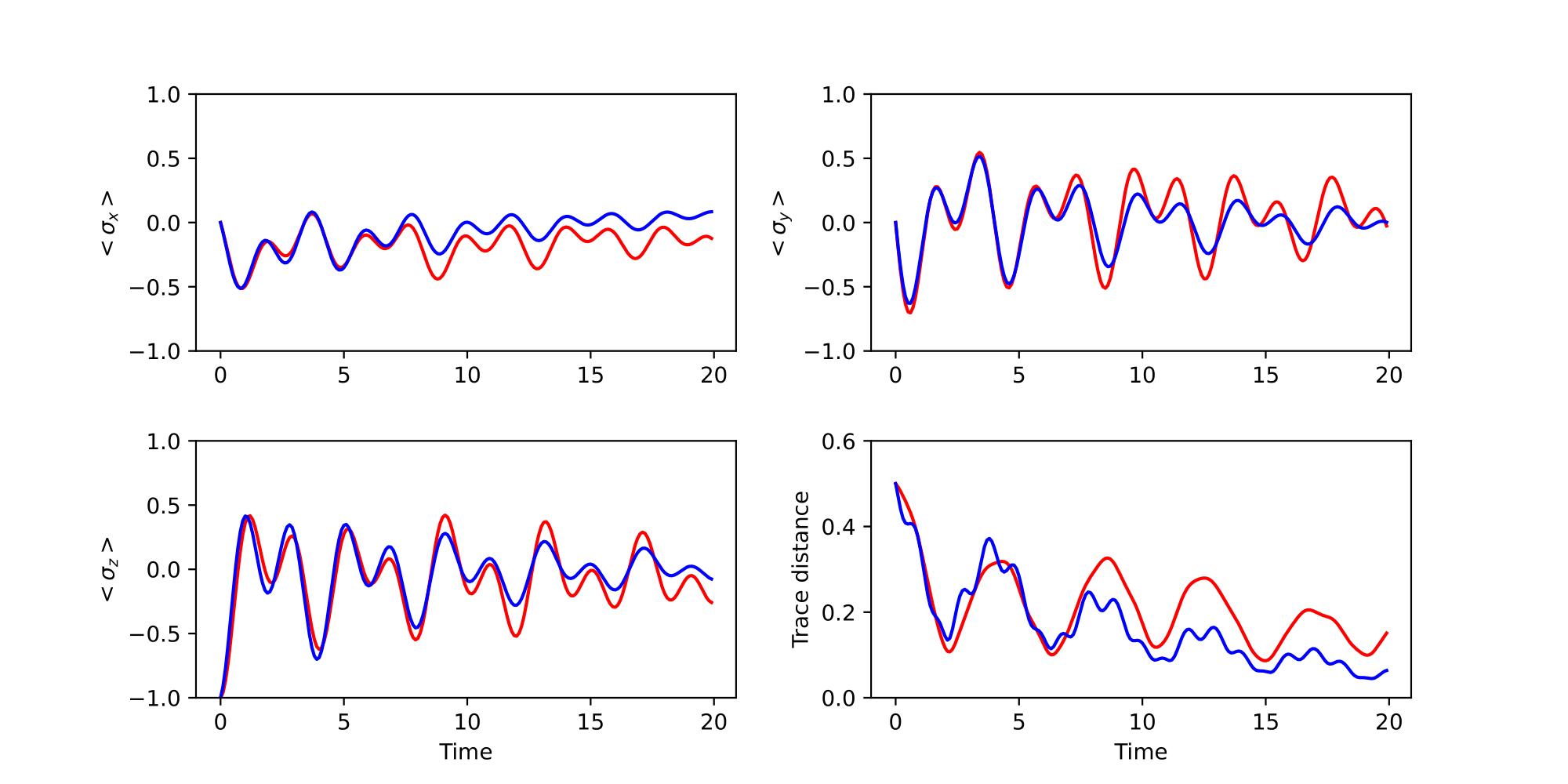} 
    \caption{Short term comparison of the real dynamics and embedded dynamics. The blue and red graphs correspond to the model and real dynamics respectively.}
    \label{fig:image1}
  \end{minipage}\hfill
  \begin{minipage}{0.45\textwidth}
    \centering
    \includegraphics[width=\linewidth]{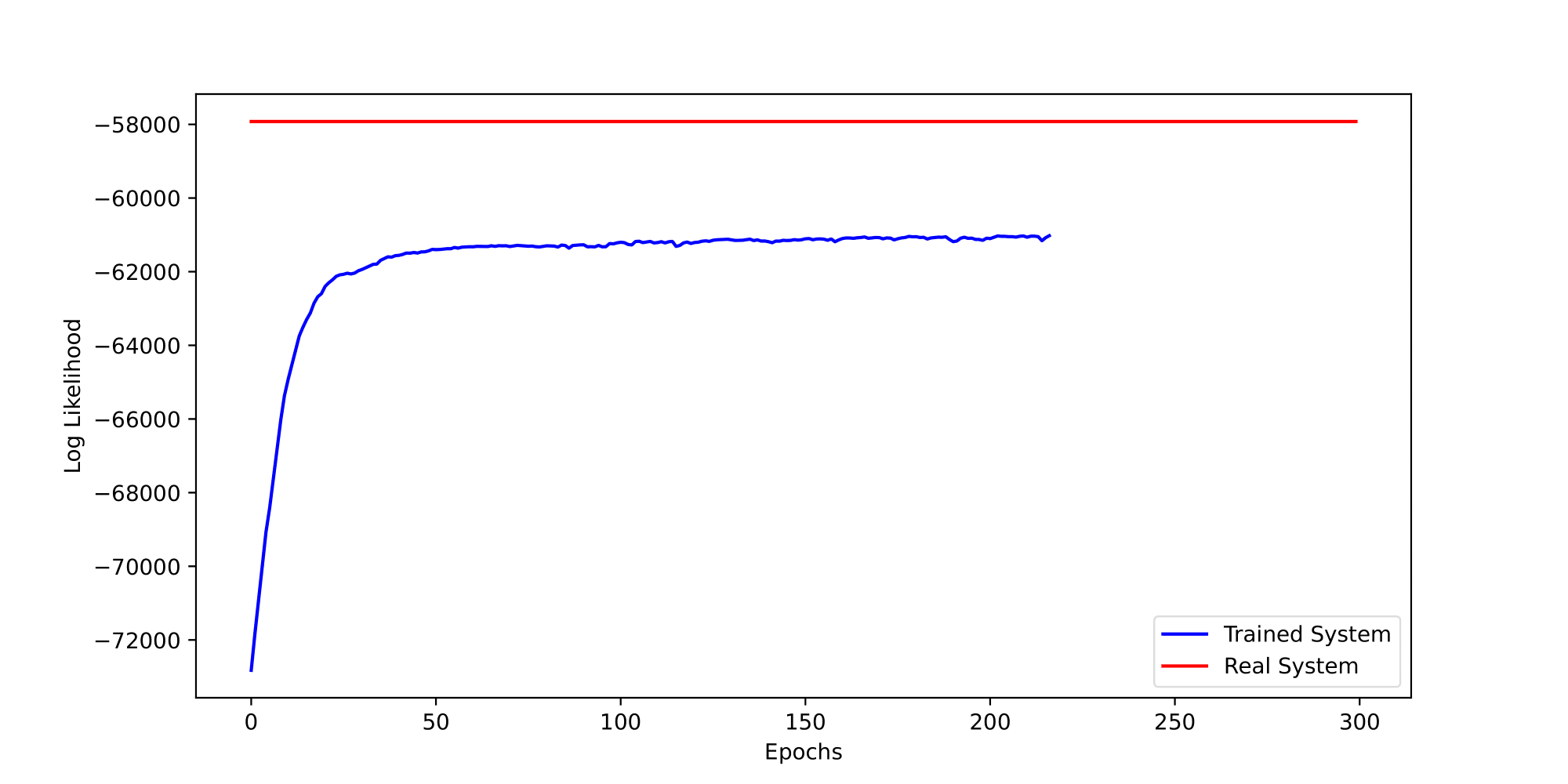} 
    \caption{Log probabilities of the model (blue) is graphed against the epoch number. The system converges early, but later gradually approaches the real system probabilities, reflecting the nonconvex nature of the optimization problem. }
    \label{fig:image2}
  \end{minipage}
\end{figure}

\begin{figure}[ht]
  \centering
  \begin{minipage}{0.45\textwidth}
    \centering
    \includegraphics[width=\linewidth]{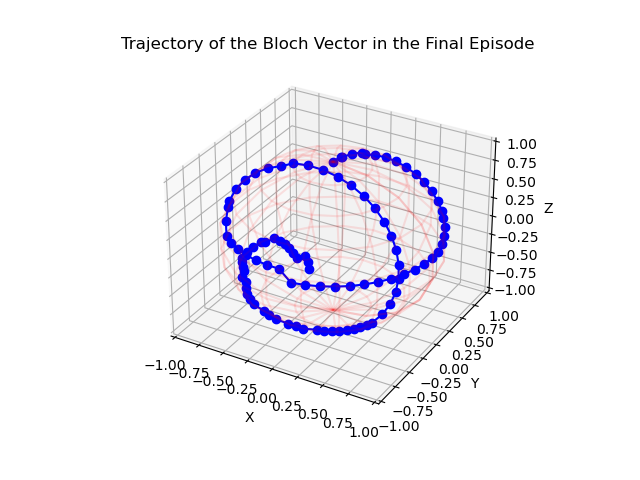} 
    \caption{The control policy of the qubit from excited state to ground state under weak coupling with QHO environment demonstrates ability to navigate control landscape in 50 control steps.}
    \label{fig:image3}
  \end{minipage}
\end{figure}

The model learning phase demonstrated rapid convergence during the initial epochs, with significant improvements up to epoch 50. Despite a decrease in the rate of convergence, the model continued to enhance its accuracy gradually. This trend underscores the model's capacity to adapt and refine its predictions over time, even in the face of the complex, nonconvex optimization landscape inherent in quantum dynamics modeling.

The decision to truncate the Fock state number of the Quantum Harmonic Oscillator (QHO) to \( N=5 \) was driven by computational efficiency considerations. This truncation proved to be most effective when the embedded dimension \( d_{ER} \) was initialized at 5, balancing the model's fidelity against computational constraints. This balance was crucial for managing fluctuations in the Bloch vector, especially for smaller values of \( d_{ER} \), which tended to average out these fluctuations.

For short durations, the model exhibited high efficacy, closely mirroring the real dynamics. However, for time scales significantly exceeding the duration of the dataset (\( \Delta t \times 10^5 \) measurements), discrepancies between the model and actual dynamics became more pronounced. These deviations suggest that for long-term predictions, a more sophisticated approach to modeling relaxation dynamics might be warranted, one as the embedded dimension \( d_{ER} \) increases. While further parameterized model could potentially offer greater accuracy, the trade-off in terms of training speed and computational demand necessitates careful consideration.

In the context of policy learning, the agent demonstrated proficiency in navigating states under conditions of low system-environment coupling. However, with an increase in the coupling constant, the purity of the state was compromised, indicating challenges in mitigating decoherence and reversing relaxation through non-unital open quantum system dynamics. This suggests that while the current approach is effective to a degree, exploring strategies such as model predictive control might offer a more robust solution for actively managing decoherence, albeit at the risk of overfitting to the learned model.

The initial success of integrating these methodologies lays a promising foundation for future research. It paves the way for enhancing the accuracy and scalability of models dealing with larger systems and more intricate bath dynamics. The insights gained from this study not only contribute to the theoretical understanding of quantum systems but also hold potential for practical applications in quantum computing, where managing decoherence is a critical challenge.

In conclusion, this study has demonstrated the feasibility and potential of using truncated models and policy gradient methods for simulating and controlling quantum systems. The ability to accurately predict and influence quantum system dynamics, even within computational constraints, opens new avenues for advancing quantum technology applications.

\bibliography{references}

\end{document}